\pdfoutput=1

\documentclass[11pt,twoside,a4paper,cmspaper,final,collab]{cms-tdr}
\def\svnVersion{487086P}\def\cmsCernNoTag{CERN-EP-2018-277}\def\cmsCernDate{\today}\def\cmsMessage{Published in the Journal of High Energy Physics as \href{http://dx.doi.org/10.1007/JHEP01(2019)154}{\doi{10.1007/JHEP01(2019)154.}}}

\begin{document}\cmsNoteHeader{SUS-17-012}

\hyphenation{had-ron-i-za-tion}
\hyphenation{cal-or-i-me-ter}
\hyphenation{de-vices}
\RCS$HeadURL: svn+ssh://svn.cern.ch/reps/tdr2/papers/SUS-17-012/trunk/SUS-17-012.tex $
\RCS$Id: SUS-17-012.tex 487084 2019-01-23 23:31:03Z msun $

\providecommand{\NA}{\ensuremath{\text{---}}}

\cmsNoteHeader{SUS-17-012}

\title{Search for supersymmetry in events with a photon, a lepton, and
    missing transverse momentum in proton-proton collisions at
    $\sqrt{s} = 13\TeV$}

\date{\today}

\abstract{Results of a search for supersymmetry are presented using
  events with a photon, an electron or muon, and large missing
  transverse momentum. The analysis is based on a data sample
  corresponding to an integrated luminosity of 35.9\fbinv of
  proton-proton collisions at $\sqrt{s} = 13\TeV$, produced by the LHC
  and collected with the CMS detector in 2016. Theoretical models with
  gauge-mediated supersymmetry breaking predict events with photons in
  the final state, as well as electroweak gauge bosons decaying to
  leptons. Searches for events with a photon, a lepton, and missing
  transverse momentum are sensitive probes of these models.  No excess
  of events is observed beyond expectations from standard model
  processes. The results of the search are interpreted in the context of
  simplified models inspired by gauge-mediated supersymmetry
  breaking. These models are used to derive upper limits on the
  production cross sections and set lower bounds on masses of
  supersymmetric particles. Gaugino masses below 930\GeV are excluded at
  the 95\% confidence level in a simplified model with electroweak
  production of a neutralino and chargino.  For simplified models of
  gluino and squark pair production, gluino masses up to 1.75\TeV and
  squark masses up to 1.43\TeV are excluded at the 95\% confidence
  level.}

\hypersetup{
pdfauthor={CMS Collaboration},
pdftitle={Search for supersymmetry in events with a photon, a lepton, and
    missing transverse momentum in proton-proton collisions at
    sqrt(s) = 13 TeV},
pdfsubject={CMS},
pdfkeywords={CMS, physics, SUSY}}

\maketitle

\section{Introduction}
\label{sec:introduction}

The search for supersymmetry (SUSY), a popular extension of the standard
model (SM) of particle physics, is a central piece of the physics
program at the CERN LHC. Models utilizing a general gauge-mediated (GGM)
SUSY mechanism~\cite{Dine:1981gu, AlvarezGaume:1981wy, Nappi:1982hm,
  Dine:1993yw, Dine:1994vc, Dine:1995ag}, with the assumption that {R
  parity}~\cite{Farrar:1978xj} is conserved, often lead to final states
containing photons and significant transverse momentum
imbalance~\cite{Dimopoulos:1996gy, Martin:1996zb, Poppitz:1996xw,
  Meade:2008wd, Buican:2008ws, Abel:2009ve, Carpenter:2008wi,
  Dumitrescu:2010ha}. Final states with an additional lepton enhance the
sensitivity to the electroweak (EW) production of SUSY particles, making
signatures with both leptons and photons an important part of the SUSY
search program at the LHC.

In GGM SUSY models, the lightest SUSY particle (LSP), taken to be the
gravitino \PXXSG, is both stable and weakly interacting. It escapes
detection, leading to missing momentum in the event. Except for direct
LSP pair production, each produced SUSY particle initiates a decay chain
that yields the next-to-lightest SUSY particle (NLSP) decaying to the
LSP. The signature of the event depends sensitively on the nature of the
NLSP. In most GGM models, the NLSP is taken to be a bino- or wino-like
neutralino or a wino-like chargino, where the bino and wino are the
superpartners of the SM U(1) and SU(2) gauge particles,
respectively. Typically, a neutral NLSP {\PSGcz} will decay to a photon
or a {\cPZ} boson, while a charged NLSP {\PSGcpm} will produce a {\PW}
boson, where both vector bosons can decay leptonically.

In this paper, the results are presented of a search for SUSY in events
with one photon \Pgg, at least one lepton $\ell$ (electron or muon), and
large transverse momentum imbalance. This signature suppresses many SM
backgrounds, avoiding the need for additional requirements such as
associated jet activity.  This makes it possible to include events with
low jet activity, increasing the sensitivity to SUSY scenarios with EW
production, in which the absence of colored SUSY particles in the decay
chain leads to lower final-state jet activity in these models.

The data sample corresponds to an integrated luminosity of 35.9\fbinv of
proton-proton ({\Pp\Pp}) collision data at $\sqrt{s} = 13\TeV$,
collected with the CMS detector at the LHC in 2016. Similar searches
with a photon plus lepton signature were conducted by the
ATLAS~\cite{Aad:2015hea} and
CMS~\cite{Chatrchyan:2011ah,Khachatryan:2015exa} experiments at
$\sqrt{s} = 7$ and 8\TeV. Searches for SUSY in GGM scenarios have also
been conducted in the single-photon~\cite{2018118,Sirunyan:2275103} and
two-photon~\cite{Aaboud:2018doq} channels at $\sqrt{s} = 13\TeV$. None
of these analyses observed any significant excess of events over their
respective SM predictions. This paper improves the sensitivity of the
previous CMS result obtained at $\sqrt{s} = 8\TeV$~\cite{CMS:2015loa}.

\begin{figure*}[hbtp]
		\centering
        \includegraphics[width=0.32\textwidth]{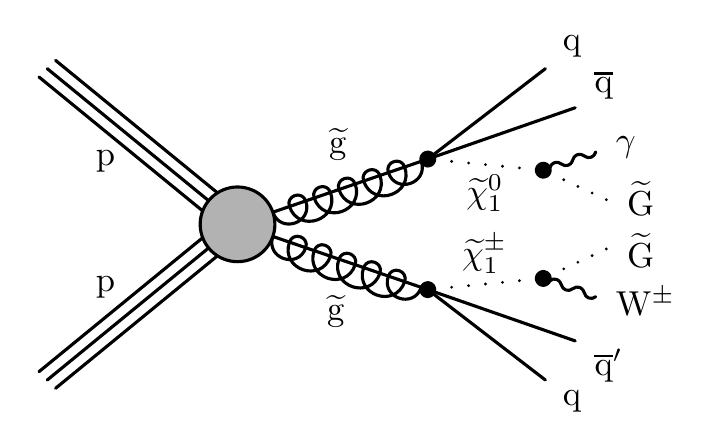}
        \includegraphics[width=0.32\textwidth]{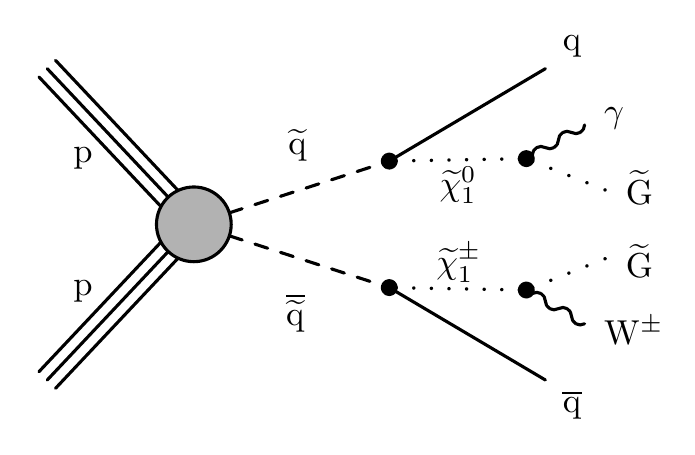}
        \includegraphics[width=0.32\textwidth]{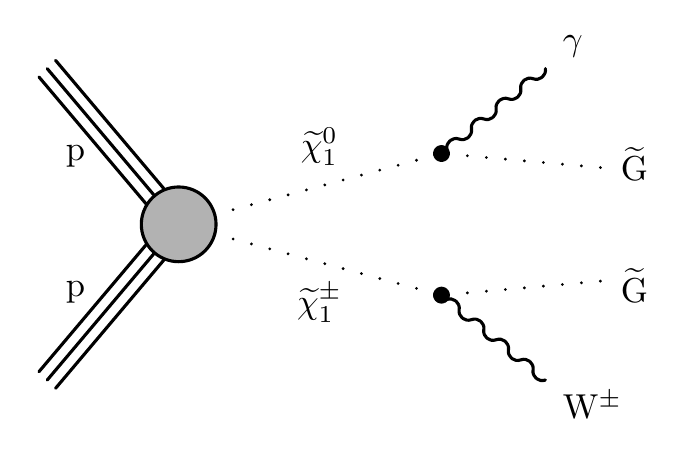}
        \caption{Diagrams showing the production and decay modes
        of the signal models T5Wg (left), T6Wg (center), and TChiWg
            (right) considered in this analysis.}
        \label{fig:Figure_002}
\end{figure*}

The diagrams in figure~\ref{fig:Figure_002} provide examples of the decays
studied in this analysis. Simplified models~\cite{Chatrchyan:2013sza}
are used for the interpretation of the results. The three simplified
models considered are denoted as T5Wg, T6Wg, and TChiWg, where T5Wg
assumes gluino ({\PSg}) pair production, T6Wg squark ({\PSq}) pair
production, and TChiWg the direct EW production of a neutralino and
chargino. For simplicity, we assume the {\PSGcz} and {\PSGcpm} are
mass-degenerate co-NLSPs and are therefore produced at equal rates. The
decay of the NLSP {\PSGcpm} (\PSGcz) produces a gravitino LSP with a
{\Wpm} ({\Pgg}). We assume a 50\% branching fraction to either the
{\PSGcz} or the {\PSGcpm} in the decays $\PSg\to \Pq\Paq\PSGcz/\PSGcpm$
and $\PSq\to \Pq\PSGcz/\PSGcpm$, and 100\% branching fractions for the
decays $\PSGcz \to \Pgg\PXXSG$ and $\PSGcpm\to \Wpm\PXXSG$.

The paper is organized as follows. In section~\ref{sec:detector}, we
describe the CMS detector used to collect the data. The data samples and
object definitions used in the analysis are described in
section~\ref{sec:objectReconstruction}, and the details of the event
selection are given in section~\ref{sec:eventSelection}. The methods for
estimating the backgrounds in the analysis are discussed in
section~\ref{sec:background}, the systematic uncertainties in
section~\ref{sec:systematics}, and the results in
section~\ref{sec:result}.  Conclusions are summarized in
section~\ref{sec:interpretation}, including our exclusion limits in the
simplified-model framework.

\section{The CMS detector}
\label{sec:detector}
The central feature of the CMS apparatus is a superconducting solenoid
with an internal diameter of 6\unit{m}, providing an axial magnetic
field of 3.8\unit{T}. Within the solenoid volume are several subdetector
systems, each composed of a cylindrical barrel closed by two endcaps. At
the core is a silicon pixel and strip tracker, providing a precise
measurement of the trajectories of charged particles. The energy of
photons and electrons is measured by a lead tungstate crystal
electromagnetic calorimeter (ECAL), covering the pseudorapidity range
$\abs{\eta} < 1.479$ in the barrel and $1.479 < \abs{\eta} < 3.0$ in the
endcap. Surrounding the ECAL is a brass and scintillator sampling hadron
calorimeter~(HCAL) with $\abs{\eta} < 3.0$ coverage. Forward
calorimeters extend the calorimeter coverage up to $\abs{\eta} =
5.0$. Muons are measured in gas-ionization detectors embedded in the
steel flux-return yoke outside the solenoid.

In the barrel section of the ECAL, an energy resolution of about 1\% is
achieved for unconverted and late-converting photons with transverse
momentum $\PT \approx 10\GeV$. The remaining barrel photons have a
resolution of about 1.3\% up to $\abs{\eta} < 1.0$, rising to about
2.5\% for $\abs{\eta} = 1.4$~\cite{Khachatryan:2015iwa}.

The electron momentum is estimated by combining the energy measurement
in the ECAL with the momentum measurement in the tracker. The momentum
resolution for electrons with $\pt\approx 45\GeV$ from $\cPZ \to
\Pep\Pem$ decays ranges from 1.7 to 4.5\%.  It is generally better in
the barrel region than in the endcaps, and also depends on the
bremsstrahlung energy emitted by the electron as it traverses the
material in front of the ECAL~\cite{Khachatryan:2015hwa}.

Muons are measured in the range $\abs{\eta} < 2.4$, with detector
elements based on three technologies: drift tubes, cathode strip
chambers, and resistive plate chambers. Matching muons to tracks
reconstructed in the silicon tracker results in a relative transverse
momentum resolution, for muons with \pt up to 100\GeV, of 1\% in the
barrel and 3\% in the endcaps.  The \pt resolution in the barrel is
better than 7\% for muons with \pt up to 1\TeV~\cite{Sirunyan:2018fpa}.

A detailed description of the CMS detector, together with the definition
of the coordinate system used and the relevant kinematic variables, can
be found in Ref.~\cite{CMS:detector}.

\section{Object reconstruction and simulated samples}
\label{sec:objectReconstruction}

Physics objects are defined using the particle-flow (PF)
algorithm~\cite{Sirunyan:2017ulk}, which aims to reconstruct and
identify each individual particle in an event via an optimized
combination of information from different elements of the CMS
detector. The PF candidates are classified as photons, charged hadrons,
neutral hadrons, electrons, or muons. The PF method also allows the
identification and mitigation of particles from additional {\Pp\Pp}
interactions in the same or adjacent beam crossings (pileup).

Photons are reconstructed from clusters of energy deposits in the
ECAL. To distinguish photon candidates from electrons, photon objects
are rejected if a matching pixel detector track segment from the silicon
tracker is identified.  Photon candidates used in this analysis are
identified with a set of loose quality criteria with an average
selection efficiency of 90\%. We require such photon candidates to be
associated with an energy deposit in the HCAL having no more than 6\% of
the energy deposited in the ECAL, and a shower shape in the $\eta$
direction consistent with that of a genuine photon.  In addition, the
photons are required to have more than 50\% of their cluster energy
deposited in the $3{\times}3$ array of crystals centered on the most
energetic crystal.

To further suppress the misidentification of hadrons as photons, a
PF-based isolation requirement is imposed. The isolation variable is
calculated by summing the magnitude of the transverse momentum of all PF
charged hadrons, neutral hadrons, and other photons within a cone of
$\Delta R \equiv \sqrt{\smash[b]{(\Delta\eta)^2 + (\Delta\phi)^2}} =$
0.3, where $\phi$ is the azimuthal angle in radians, around the
candidate photon direction. We required this variable not to exceed
fixed values that are set to achieve a desirable balance between
identification efficiency and misidentification rate. The photon object
that is being identified is not included in the isolation sums, and
charged hadrons are included only if they are associated with the
primary {\Pp\Pp} interaction vertex.  The reconstructed vertex with the
largest value of summed physics-object $\pt^2$ is taken to be the
primary $\Pp\Pp$ vertex. The physics objects are the jets, clustered
using the jet-finding algorithm~\cite{CMS:ak1,CMS:ak2} with the tracks
assigned to the vertex as inputs, and the associated missing transverse
momentum, taken as the negative vector \pt sum of those jets.

Electrons are found by associating tracks reconstructed in the silicon
tracker with ECAL clusters. The electron candidates are required to be
within the fiducial region of $\abs{\eta} < 2.5$, where the tracker
coverage ends. Identification of electrons is based on the shower shape
of the ECAL cluster, the HCAL-to-ECAL energy ratio, the geometric
matching between the cluster and the track, the quality of the track
reconstruction, and the isolation variable. To enhance the
identification efficiency, the isolation variable is calculated from the
transverse momenta of photons, charged hadrons, and neutral hadrons
within a $\Delta{R}$ cone whose radius is variable depending on the
electron \pt~\cite{Rehermann2011}, and which is also corrected for the
effects of pileup~\cite{Cacciari:2007fd}.

The reconstruction of muons is based on associating tracks from the
silicon tracker with those in the muon system. A set of muon
identification criteria, based on the goodness of the track fit and the
quality of muon reconstruction, is applied to select the muon
candidates, having an efficiency greater than 98\% for genuine
muons~\cite{Sirunyan:2018fpa}. Muons are also required to be isolated
from other objects in the event using a similar isolation
variable~\cite{Sirunyan:2018fpa} as in the electron identification.

Jets are reconstructed starting with all PF candidates that are
clustered using the anti-\kt algorithm~\cite{CMS:ak1,CMS:ak2} with a
distance parameter that determines the nominal jet radius of
$R=0.4$. The jet energies are corrected for detector response, as well
as an offset energy from pileup interactions~\cite{Cacciari:2007fd}. Jet
candidates considered in this analysis are required to have $\pt>30\GeV$
and be within the $\abs{\eta} < 2.5$ region. Tracks associated with the
jet are required to be consistent with originating from the primary
vertex. The missing transverse momentum vector \ptvecmiss is given by
the negative vector \pt sum of all PF objects, with jet energy
corrections~\cite{Cacciari:2007fd,Chatrchyan:2011ds} applied. The
magnitude of \ptvecmiss is referred to as the missing transverse
momentum \ptmiss. The near hermiticity of the CMS detector allows for
accurate measurements of \ptmiss. Dedicated filters are applied to
remove events with \ptmiss induced by beam halo, noise in the detector,
or poorly reconstructed muons~\cite{1748-0221-10-02-P02006}.

Monte Carlo (MC) simulated events are used to model the SM backgrounds,
validate the background estimation methods, and study the SUSY signal
yields. In order to study the SM backgrounds, discussed more fully in
section~\ref{sec:background}, samples of {\PW\Pgg} events are generated
with \MGvATNLO 2.3.3~\cite{CMS:madgraph} at leading order (LO), while
the \cPZ\Pgg, Drell--Yan, \PW\PW($+ \Pgg$), \PW\cPZ($+ \Pgg$), and
\ttbar($+ \Pgg$) background processes are generated at next-to-leading
order (NLO). All samples use the NNPDF 3.0~\cite{Ball2015} parton
distribution functions (PDFs). The generated events are interfaced with
\PYTHIA 8.205 or 8.212~\cite{CMS:pythia} with the CUETP8M1 underlying
event tune~\cite{CMS:tune} for simulation of parton showering and
hadronization. Renormalization and factorization scales and PDF
uncertainties are derived with the use of the SysCalc
package~\cite{Kalogeropoulos:2018cke}.  The \cPZ\Pgg, Drell--Yan,
\PW\PW($+ \Pgg$), \PW\cPZ($+ \Pgg$), and \ttbar($+ \Pgg$) samples are
scaled to the integrated luminosity using the theoretical cross sections
at NLO precision~\cite{CMS:madgraph}. For the {\PW\Pgg} sample, a
next-to-NLO (NNLO) scale factor of 1.34~\cite{THEORY:kfactor} is applied
to the LO cross section to account for higher-order corrections. The CMS
detector response is simulated using a \GEANTfour-based~\cite{EXP:geant}
package. The effects of pileup are modeled in the simulation by
overlaying simulated minimum-bias events on the corresponding
hard-scattering event, and the distribution of the pileup vertices is
reweighted to match that observed in data.

The signal events in the three simplified models introduced in
section~\ref{sec:introduction} are generated with \MGvATNLO at LO. The
cross sections are calculated at NLO plus next-to-leading-logarithm
(NLL) accuracy~\cite{Beenakker:1996ch,Beenakker:1999xh,Fuks:2012qx,Fuks:2013vua,Borschensky2014}.  The
generated events are processed with a fast simulation of the CMS
detector response~\cite{CMS:2010spa}. Scale factors are applied to
compensate for any differences with respect to the full simulation.

To improve the \MGvATNLO modeling of initial-state radiation (ISR),
which affects the total transverse momentum of the event, the ISR
transverse momentum ($\pt^{\textrm{ISR}}$) distributions of the MC
{\PW\Pgg} and {\cPZ\Pgg} events are weighted to agree with those in
data. This reweighting procedure is based on studies of the transverse
momentum of {\cPZ} boson events~\cite{CMS:isrweight}. The reweighting
factors range from 1.11 for $\pt^{\textrm{ISR}} \approx 125\GeV$ to 0.64
for $\pt^{\textrm{ISR}} > 300\GeV$. We take the deviation of the
reweighting factors from 1.0 as an estimate of the systematic
uncertainty in the reweighting procedure.

\section{Event selection}
\label{sec:eventSelection}

The analysis is performed in both the {\re\Pgg} and $\mu${\Pgg}
channels. The {\re\Pgg} data sample is collected using a diphoton
trigger~\cite{CMS:trigger} requiring at least two isolated
electromagnetic objects with \pt thresholds of 30 and 18\GeV for the
highest \pt and second-highest-\pt electromagnetic object, respectively,
that satisfy loose identification criteria and have an invariant mass
$M_{\Pgg\Pgg} > 90\GeV$.  The trigger does not veto photon objects that
can be matched to a track from the silicon tracker, allowing events with
a photon and an electron to also pass the trigger selections.  The
$\mu${\Pgg} events are collected using a combination of two muon+photon
triggers, one requiring the presence of an isolated photon with $\pt >
30\GeV$ and a muon with $\pt > 17\GeV$, and the other using symmetric
\pt thresholds of 38{\GeV} for both objects, with no photon isolation
criteria.  With the selection criteria described below, the average
trigger efficiency for the investigated SUSY signal models is found to
be 96\% for $\Pe${\Pgg} and 94\% for $\mu$\Pgg.

Candidate signal events are required to contain at least one isolated
photon with $\pt^{\Pgg} > 35\GeV$ and $\abs{\eta} < 1.44$ and at least
one isolated electron (muon) with $\pt > 25\GeV$ and $\abs{\eta} < 2.5$
(2.4). To ensure a high reconstruction efficiency, electrons in the
barrel-endcap transition region $1.44 < \abs{\eta} < 1.56$ are rejected.
If more than one electron (muon) satisfies the selection criteria, the
highest \pt candidate is selected. To suppress events with photons from
final-state radiation, photon candidates are vetoed if they are within
$\Delta R < 0.3$ of any reconstructed electron or muon. In addition, the
highest \pt photon is required to be separated from the highest \pt
lepton by $\Delta R>0.8$.  In the {\re\Pgg} channel, the {\re\Pgg}
invariant mass must be at least 10\GeV greater than the world-average
{\cPZ} boson mass~\cite{EXP:zmass} to reduce the contribution of $\cPZ
\to \re^+\re^-$ events, where one of the electrons is misidentified as a
photon.

For each event we compute the transverse mass \mT of the lepton plus
\ptmiss system to help discriminate between the SUSY signal and SM
backgrounds.  The quantity \mT is defined as $\mT =
\sqrt{\smash[b]{2\pt^{\ell}\ptmiss[1-\cos(\Delta\phi(\ell,\ptvecmiss))]}}$,
where $\pt^{\ell}$ is the magnitude of the lepton transverse momentum
and $\Delta\phi$ is the difference in azimuthal angle between the
direction of the lepton and \ptvecmiss.  The signal region is defined as
$\ptmiss > 120\GeV$ and $\mT > 100\GeV$. Models with strongly produced
SUSY particles lead to final states with significant hadronic activity
in the form of jets. To provide additional sensitivity to these models,
we define the variable \HT as the scalar sum of the transverse momenta
of all jets that are separated from both the candidate photon and
candidate lepton by $\Delta R > 0.4$.  The signal region is later
divided into search regions as a function of \ptmiss, $\pt^{\Pgg}$, and
\HT.

\section{Background estimation}
\label{sec:background}

The SM backgrounds of events with one lepton, one photon, and
substantial \ptmiss in the final state mainly arise from three
sources. The first consists of events without a directly produced
(prompt) photon. This includes events with a photon that does not
originate from the hard-scattering event vertex, but from a nearby
pileup vertex, as well as events with an object such as an electron or
an electromagnetically rich jet that is misidentified as a photon. The
second source of background consists of events that do not contain a
prompt lepton. These typically result from the misidentification of a
jet as a lepton, or from a jet caused by the hadronization of a
heavy-flavor quark, which produces a lepton via the semileptonic decay
of the corresponding heavy-flavor meson or baryon. The final
contribution to the background comes from EW processes, primarily
{\PW\Pgg} and {\cPZ\Pgg} production. This category also includes rarer
processes such as \PW\PW\Pgg, \PW\cPZ\Pgg, and $\ttbar\Pgg$, referred to
in this paper as the ``rare EW'' background.

The contribution from EW processes is estimated via simulation, while
the backgrounds due to misidentified photons and leptons are estimated
from data, as described below.

\subsection{Backgrounds from misidentified photons}

Photon candidates are considered misidentified if they are not produced
directly in the hard-scattering process, or if they result from a
misidentified object. The latter constitute the majority of
misidentified photons and can occur in two cases: when a large fraction
of the energy of a jet is carried by a neutral pion decaying into two
almost collinear photons, or when an electron fails to register hits in
the pixel tracker.  In both cases, a misidentified photon is
reconstructed. Signal candidate events with misidentified photons from
jets can arise from the process $\PW(\to \ell\Pgn) +$ jets, where a
{\PGpz} or $\eta$ meson in the jet decays to photons. Signal candidate
events with misidentified photons from electrons can arise from
Drell--Yan dielectron production ($\Pq\Paq\to\Pgg^{*}\to\Pep\Pem$), as
well as \ttbar events with an electron in the final state.

The misidentified-photon background is estimated from collision data by
determining the mis\-identification rate from a control sample of
electron-like objects and applying it to events in a control region.
First, the control sample is formed by replacing the photon candidate
with a photon-like object, which is obtained by inverting some of the
photon identification criteria, while keeping the other selection
requirements identical to those for signal candidates. Second, the
misidentification rate is defined as the ratio of the number of
misidentified photons to the total number of photon-like objects in the
control sample. The misidentification rate is applied in a control
region, defined by $\ptmiss < 70\GeV$, to estimate the number of
misidentified photons in the control region. This estimate is then
extrapolated to the signal region.

\begin{figure}[bp]
    \centering
    \includegraphics[width=0.49\textwidth]{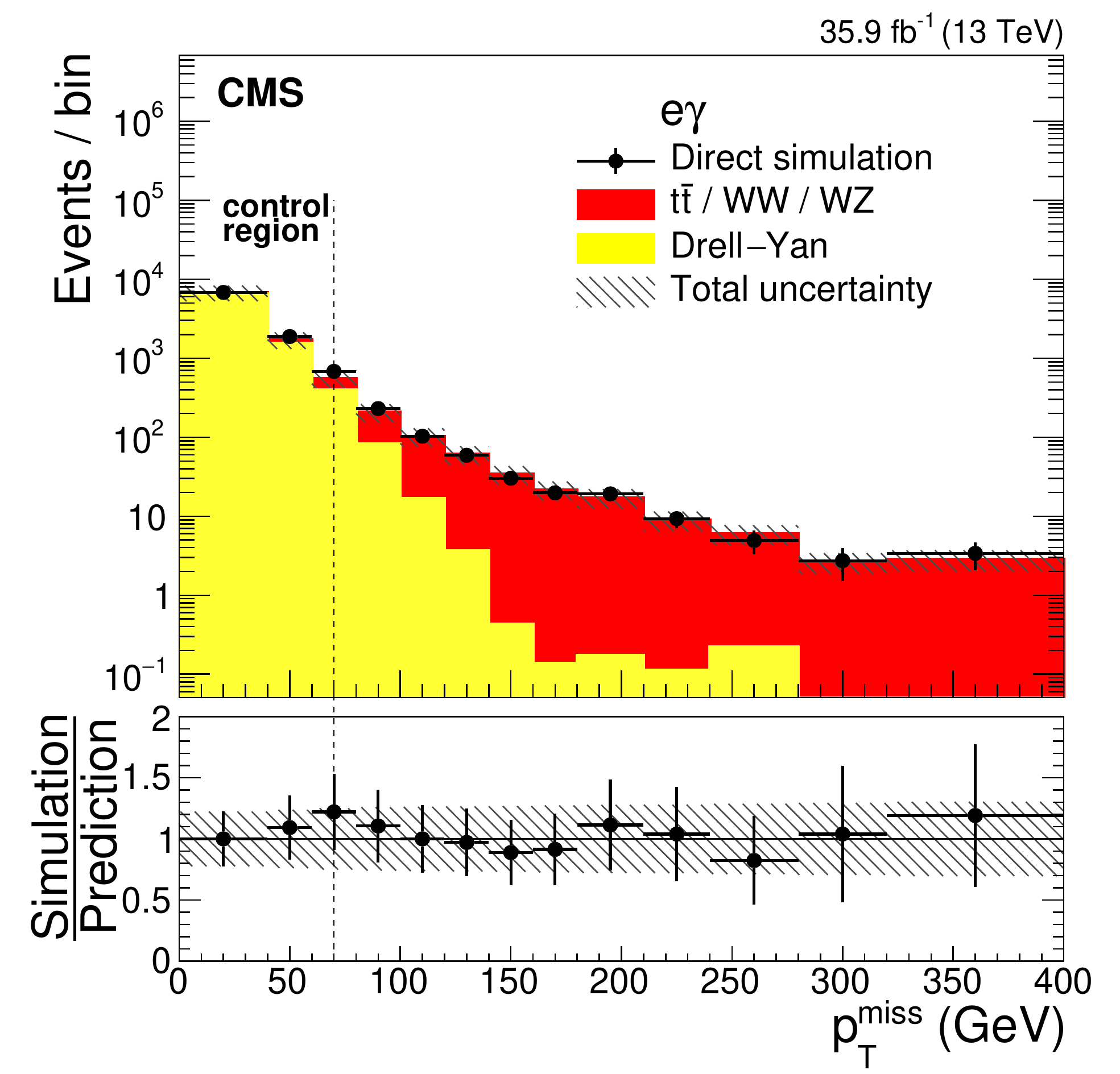}
    \includegraphics[width=0.49\textwidth]{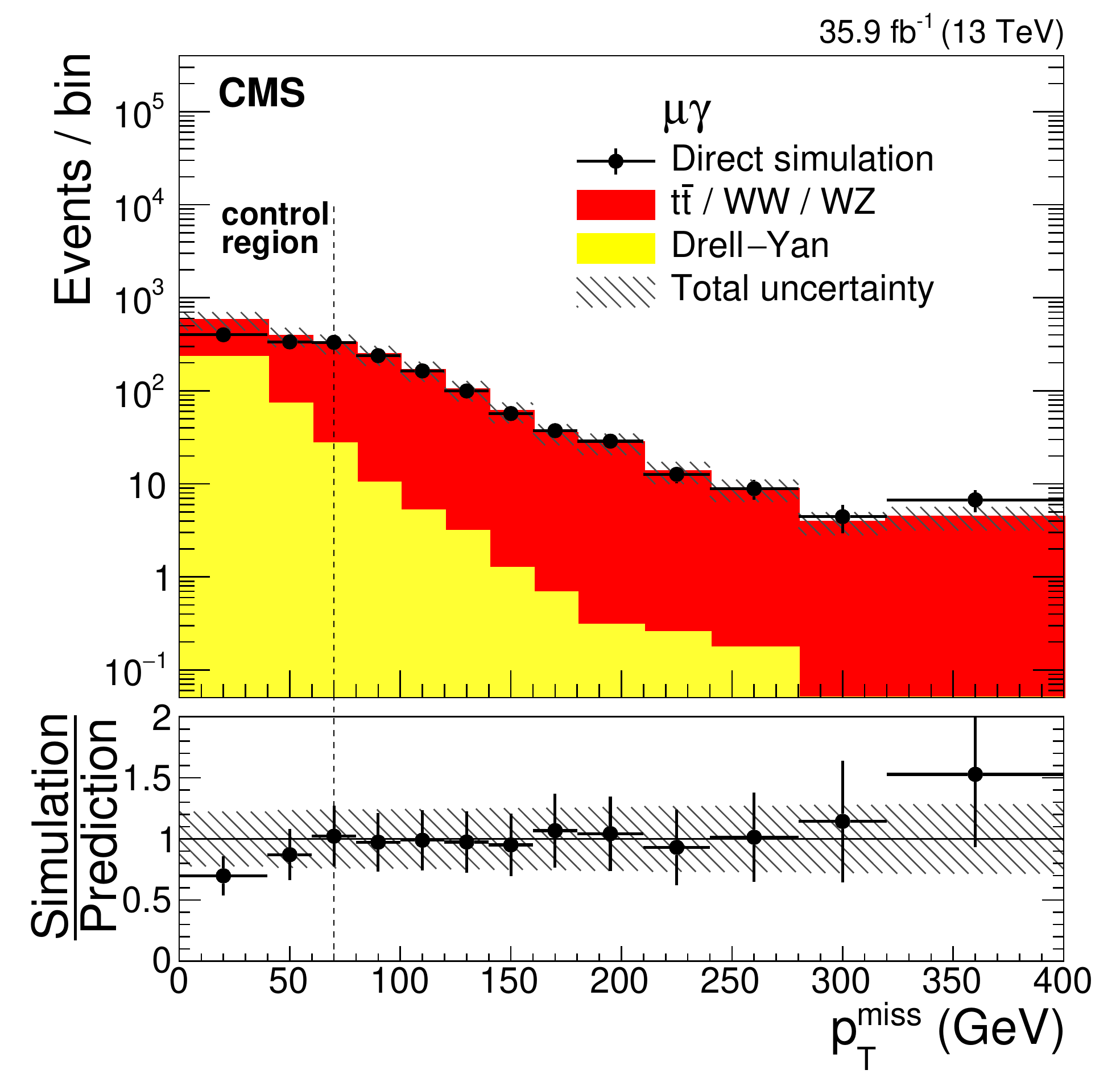}
    \caption{Verification of the \re-to-{\Pgg} misidentification
      estimation method using simulated data. The \ptmiss distribution
      for events with misidentified photons in the {\re\Pgg} (left) and
      $\mu\Pgg$ (right) channels from prediction using the control
      sample estimation method (histograms) and direct simulation
      (points), as obtained from the generator-level information of the
      simulated data. The vertical bars on the points show the
      statistical uncertainty in the simulation, while the horizontal
      bars give the bin widths. The dashed vertical line shows the
      boundary between the control and signal regions. The lower panels
      show the ratio of the predictions from direct simulation to those
      estimated with control samples. The hatched areas give
      the quadrature sum of the statistical and systematic uncertainties
      in the simulated background.  }
    \label{fig:metvalidation}
\end{figure}

Electron control samples are constructed by requiring a candidate photon
to either be associated with a seed track in the pixel detector or be
geometrically matched to a reconstructed electron within $\Delta R <
0.03$. The misidentification rate is estimated using the
``tag-and-probe'' method~\cite{CMS:tagandprobe} on a sample of $\cPZ \to
\re^+\re^-$ events in data. The rate is derived in bins of three
variables: the \pt and $\abs{\eta}$ of the probe objects, and the number
of vertices in the event $N_{\textrm{vtx}}$.  Parameterized functions
are used to model the dependence of the misidentification rate on \pt
and $N_{\textrm{vtx}}$, and binned values are used for the $\abs{\eta}$
dependence. The measured misidentification rate varies from 2.3\% for
$\pt = 35\GeV$ to 1.2\% for $\pt > 180\GeV$.  These misidentification
rates are then applied on an event-by-event basis in the control region
when estimating the misidentified-lepton backgrounds later in the signal
region.  To verify the correctness of this background estimation method,
it is tested on simulated Drell--Yan and
\ttbar/\PW\PW/\PW\cPZ~events. As shown in figure~\ref{fig:metvalidation},
good agreement is achieved in the \ptmiss distribution of these
simulated background events found using the control sample \re-to-{\Pgg}
misidentification estimation method and that found directly from the
generator-level truth information.

To estimate the jet-to-photon misidentification background, a hadronic
control sample is constructed by inverting one of the variables
characterizing the ECAL cluster shape ($\sigma_{\eta \eta}$ in
Ref.~\cite{Khachatryan:2015hwa}) and the isolation variable
requirement. The misidentification rate for the hadronic control sample
is determined through an assessment of the fraction of events with
jet-to-photon misidentification among the photon candidates. This
fraction is denoted as the ``hadron fraction''. The measurement is
performed in the control region $\ptmiss < 70\GeV$ from a fit to the
isolation variable distribution based on two templates, one representing
pure photons obtained from $\Pgg +$jet simulated MC events and one
modeling the events with jet-to-photon misidentification, where the
template for those events is obtained by inverting the $\sigma_{\eta
  \eta}$ requirement on the signal-photon candidates.  The fit to the
isolation distribution is performed in bins of $\pt^{\Pgg}$. The
resulting hadron fraction varies from 47 to 4\% for the {\re\Pgg}
channel and 18 to 4\% for the $\mu${\Pgg} channel as $\pt^{\Pgg}$
increases. The \pt distribution of the jet-to-photon background in the
control region is obtained by multiplying the \pt distribution of the
photon candidates by the hadron fraction. To extrapolate the result to
high-\pt photons, the \pt shape of the jet-to-photon backgrounds and the
control samples are modeled with the sum of two exponential functions,
and the ratio between these two functions is used to assign
event-by-event misidentification rates in the signal region. In the
{\re\Pgg} channel, the misidentification rate varies from 28\% at $\pt =
35\GeV$ to 12\% at $\pt = 200\GeV$. In the $\mu${\Pgg} channel, it drops
from 22 to 10\% as \pt goes from 35 to 200\GeV.

\subsection{Electroweak and misidentified-lepton backgrounds}

A lepton is considered to be misidentified if it doesn't originate from
a prompt {\PW} or {\cPZ} boson decay. This includes leptons from heavy-
and light-flavor hadron decays, misidentified jets, and electrons from
photon conversions. Similar to the misidentified-photon background, the
shapes of the misidentified-lepton backgrounds are modeled by control
samples, which are formed by inverting the isolation requirement of the
lepton while keeping other requirements unchanged. For electrons, the
cluster shape and the quality of the cluster-to-track matching are also
inverted to include more hadronic objects.

The SM backgrounds in final states with a lepton, a photon, and large
\ptmiss are dominated by the production of {\PW} and {\cPZ} bosons in
association with a photon, denoted as V{\Pgg} production. In particular,
neutrinos from the {\PW} boson leptonic decay escape the detector,
producing significant \ptmiss. The shape of the \ptmiss distribution
from the V{\Pgg} background is modeled by simulation, and the
normalization factors are determined together with those of the
misidentified-lepton backgrounds, as described in the next paragraph.

\begin{figure}[tbp]
    \centering
    \includegraphics[width=0.49\textwidth]{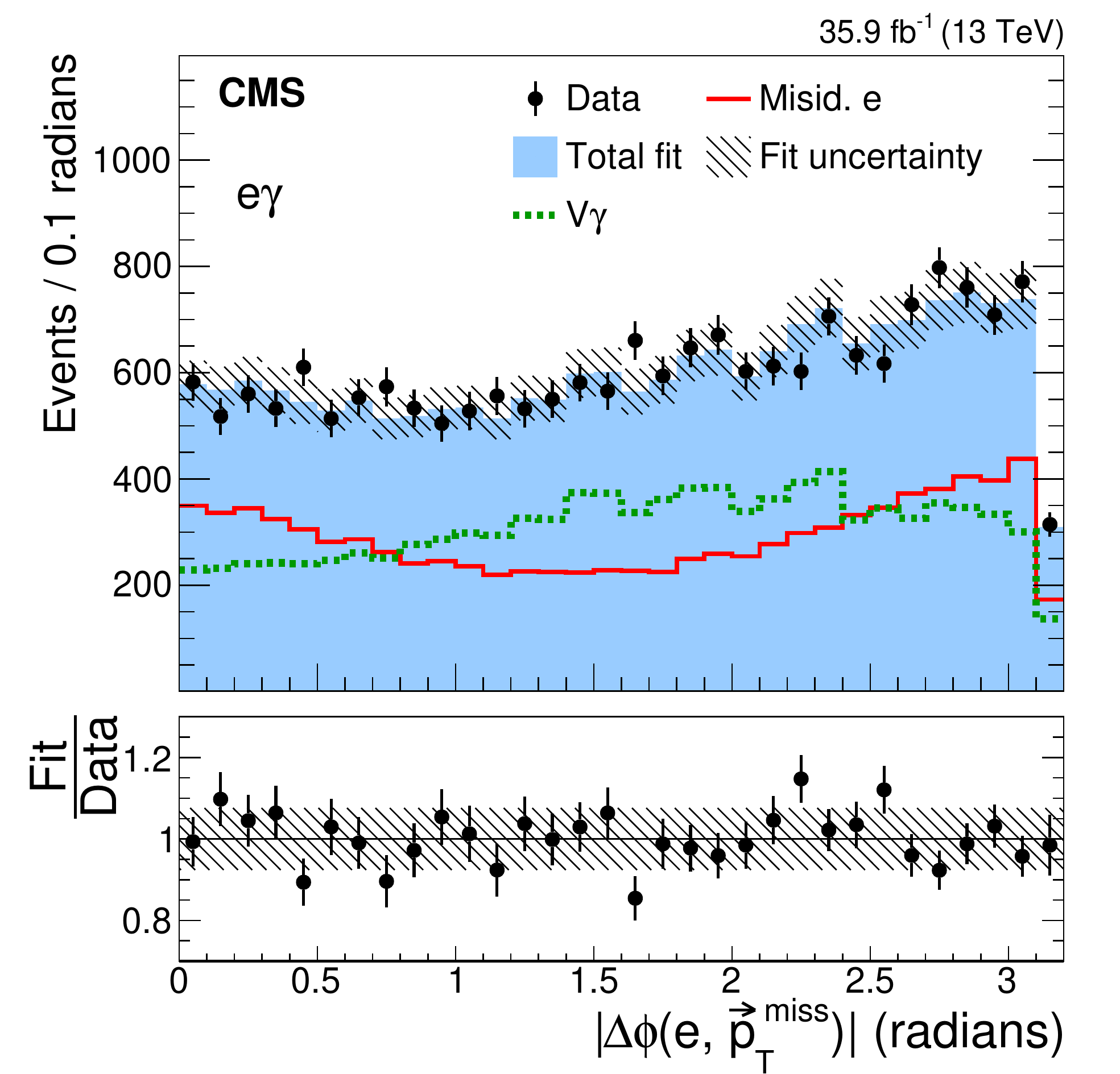}
    \includegraphics[width=0.49\textwidth]{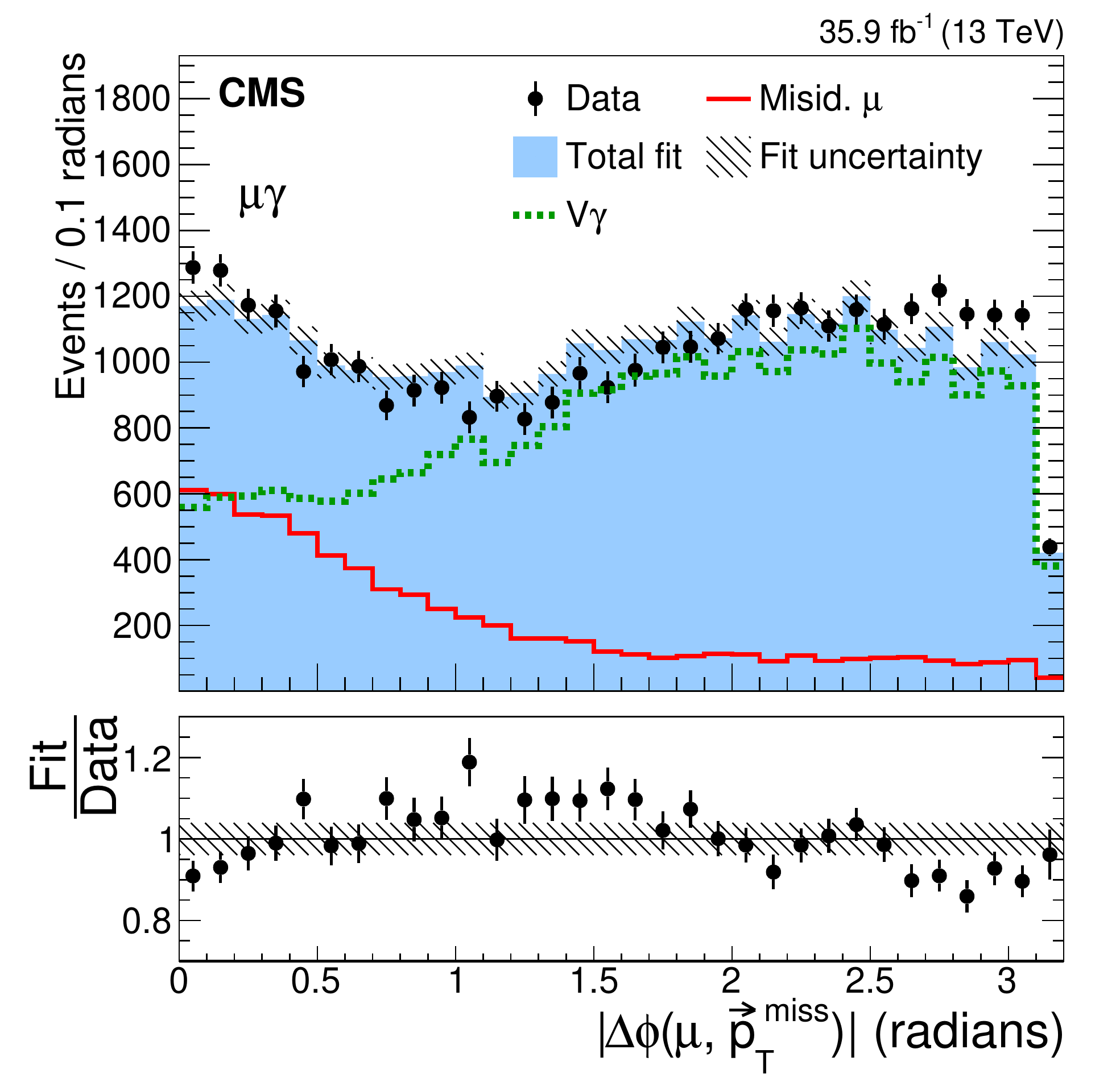}
    \caption{The $\abs{\Delta\phi(\ell,\ptvecmiss)}$ distributions for
      the data in the $40 < \ptmiss < 70\GeV$ control region (points)
      and the estimated V{\Pgg} (dashed line) and misidentified-lepton
      (solid line) backgrounds for the {\re\Pgg} (left) and $\mu${\Pgg}
      (right) channels.  The filled histogram shows the result of the
      overall fit and the hatched area indicates the fit
      uncertainty. The vertical bars on the points represent the
      statistical uncertainty in the data. The lower panels show the
      ratio of the fit result to the data.  }
    \label{fig:met2}
\end{figure}

The normalization of the V{\Pgg} and misidentified-lepton backgrounds is
determined by a two-component signal-plus-background template fit to the
distribution of $\abs{\Delta\phi(\ell,\ptvecmiss)}$, the azimuthal
angular difference between the direction of the lepton and \ptvecmiss in
the transverse plane. This fit is performed in the control region $40 <
\ptmiss < 70\GeV$, where the lower bound of 40\GeV is applied to reduce
the contribution of {\cPZ\Pgg} events.  Expected contributions from the
misidentified-photon and rare EW backgrounds such as \PW\PW($+ \Pgg$),
\PW\cPZ($+ \Pgg$), and {\ttbar}($+ \Pgg$) processes are subtracted
before the fit.  The distribution of $\abs{\Delta\phi(\ell,
  \ptvecmiss)}$ is shown in figure~\ref{fig:met2} with the fit results
overlaid.  The resulting scale factors (SFs) for the V{\Pgg} and
misidentified-lepton backgrounds in the {\re\Pgg} channel are
$SF_{\mathrm{V}\Pgg} = 1.17 \pm 0.08$ and $SF_{\mathrm{e-misid}} = 0.24
\pm 0.02$, respectively, while the SFs for the $\mu${\Pgg} channel are
$SF_{\mathrm{V}\Pgg} = 1.33 \pm 0.02 $ and $SF_{\mu-\mathrm{misid}} =
0.62 \pm 0.02$, where the uncertainties are statistical only.

\section{Systematic uncertainties}
\label{sec:systematics}

Table \ref{table:systematic} summarizes the relative systematic
uncertainties in the background estimation and signal expectation.  If
the relative uncertainties differ considerably in different kinematic
regions because of the limited number of events available for the
evaluation of the systematic uncertainties, the range of the relative
uncertainty is shown.  The main sources of systematic uncertainties are
the SFs derived from the $\abs{\Delta\phi(\ell, \ptvecmiss)}$ template
fit to the V{\Pgg} and misidentified-lepton backgrounds, and the cross
sections used to normalize the rare EW simulated samples. The systematic
uncertainty coming from the shape of the V{\Pgg} distribution is
obtained by allowing each bin of the template to vary independently
according to a Gaussian distribution. Systematic uncertainties in the
magnitude of the normalization are determined by allowing the number of
subtracted events from the estimated backgrounds to vary within their
uncertainties, as well as the PDF and renormalization and factorization
scales of the V{\Pgg} template to vary by one standard deviation around
their nominal values. For the rare EW backgrounds, a $50\%$ uncertainty
is assigned to the cross sections to cover the difference between the
calculated cross sections and the latest CMS
measurements~\cite{CMS:ttgxs,CMS:ttxs}.

The subdominant systematic uncertainties come from the modeling of the
misidentified photons. Different choices of control samples and
parameterized functions are studied to evaluate the size of these
systematic effects. The uncertainties in the number of misidentified
photons with $\pt < 200\GeV$ are less than $20\%$. A larger uncertainty,
up to 56\%, is caused by the limited number of events in the control
sample and applies only to the high-\pt bins, where the misidentified
photons contribute less than $10\%$ of the total background, resulting
in a small effect on the total background prediction. For the
backgrounds obtained from simulation, systematic uncertainties from the
jet energy scale are evaluated by varying the corresponding scale by one
standard deviation around its nominal value~\cite{Khachatryan:2016kdb}.
Uncertainties in the signal cross sections used in the simulation due to
the PDFs and the renormalization and factorization scales are taken from
Refs.~\cite{Fuks:2012qx,Fuks:2013vua,Borschensky2014}.  The additional
shape uncertainty in the signal sample due to the choice of the
renormalization and factorization scales is estimated by varying the
scales upward and downward by a factor of two with respect to their
nominal values.  Finally, the uncertainty in the integrated luminosity
of the data sample is 2.5\%~\cite{CMS-PAS-LUM-17-001}.

\begin{table}[tbph]
  \centering
  \topcaption{The relative systematic uncertainties in the SM
    background processes (third column) and the expected SUSY signal
    (fourth column). The ranges refer to the uncertainties over the
    different kinematic regions.}
  \label{table:systematic}
  \resizebox{\linewidth}{!}{
  \begin{tabular}{llcc}
  \hline
  Uncertainty source                           & Background process    & Background uncertainty (\%)  &  Signal uncertainty (\%) \\

\hline
  Jet energy scale                              & V\Pgg, rare EW        & 0--23         & 0--10              \\
  Normalization scale                           & V\Pgg, jet $\to \ell$ misid. & 20     & \NA         \\
  Cross section                                 & rare EW               & 50            & 4--37                \\
  Ident.~and trigger efficiency                 & V\Pgg, rare EW        & 1.3--6.5      & 1.3--6.5              \\
  $\Pe\to\Pgg$                          & $\Pe\to\Pgg$ misid. & 8--51                   & \NA  \\
  Jet $\to\Pgg$ shape                    & jet $\to\Pgg$ misid. & 8--56                 & \NA \\
  Misid.~lepton shape                           & jet $\to \ell$ misid.& 0--42          & \NA     \\
  ISR corrections                               &  V{\Pgg}                & 3--58         & 0--32          \\
  Integrated luminosity                         & rare EW               & 2.5           & 2.5                \\
  Pileup uncertainty                            & \NA                   & \NA           & 2--10       \\
  PDF, renormalization/factorization scales     & \NA                   & \NA           & 0--10       \\
  Fast simulation \ptmiss modeling             & \NA                   & \NA           & 0--31                     \\
\hline \\
  \end{tabular}
  }
\end{table}

\section{Results}
\label{sec:result}

\begin{figure}[tbp]
\centering
\includegraphics[width=0.39\textwidth]{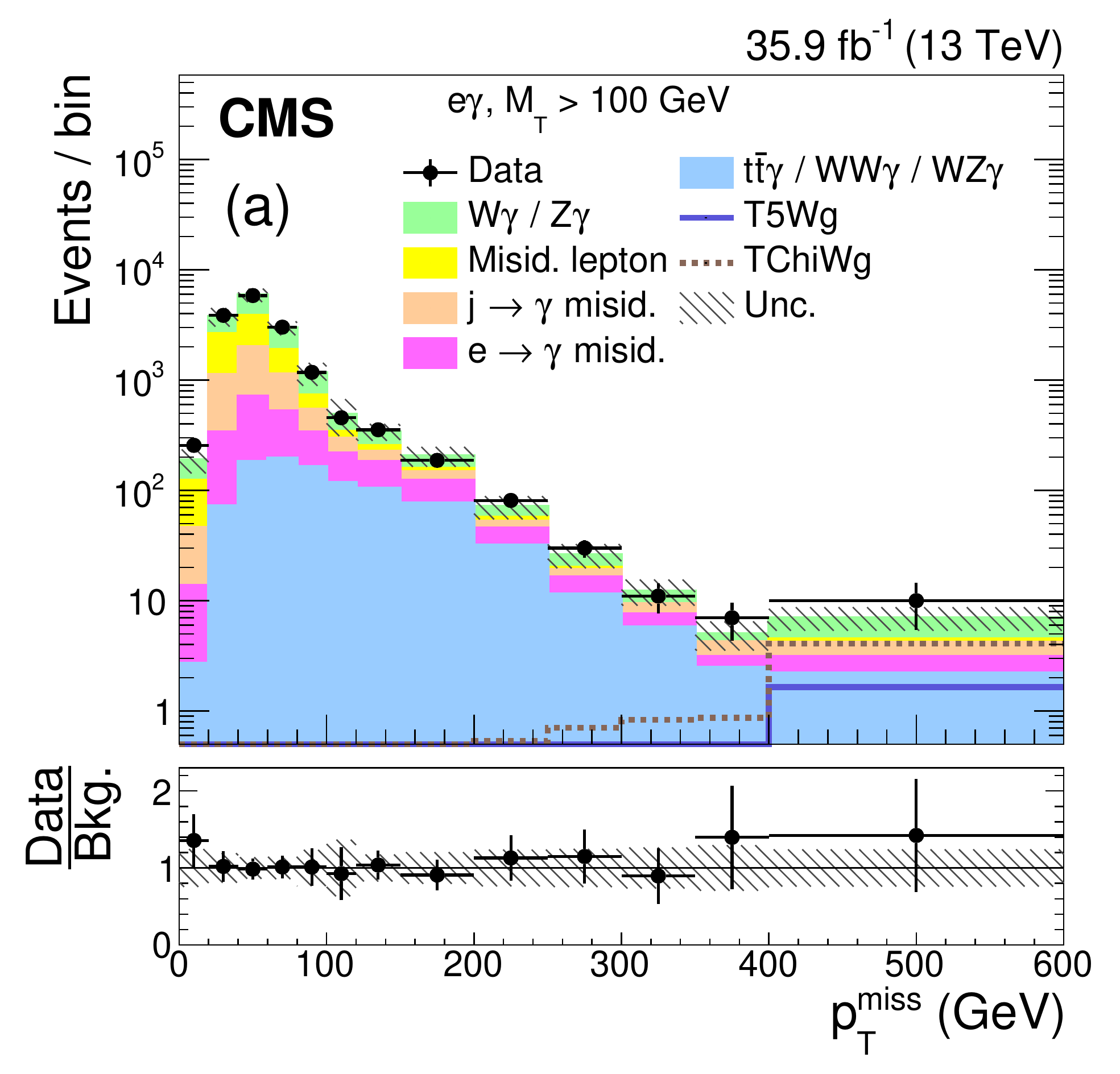}
\includegraphics[width=0.39\textwidth]{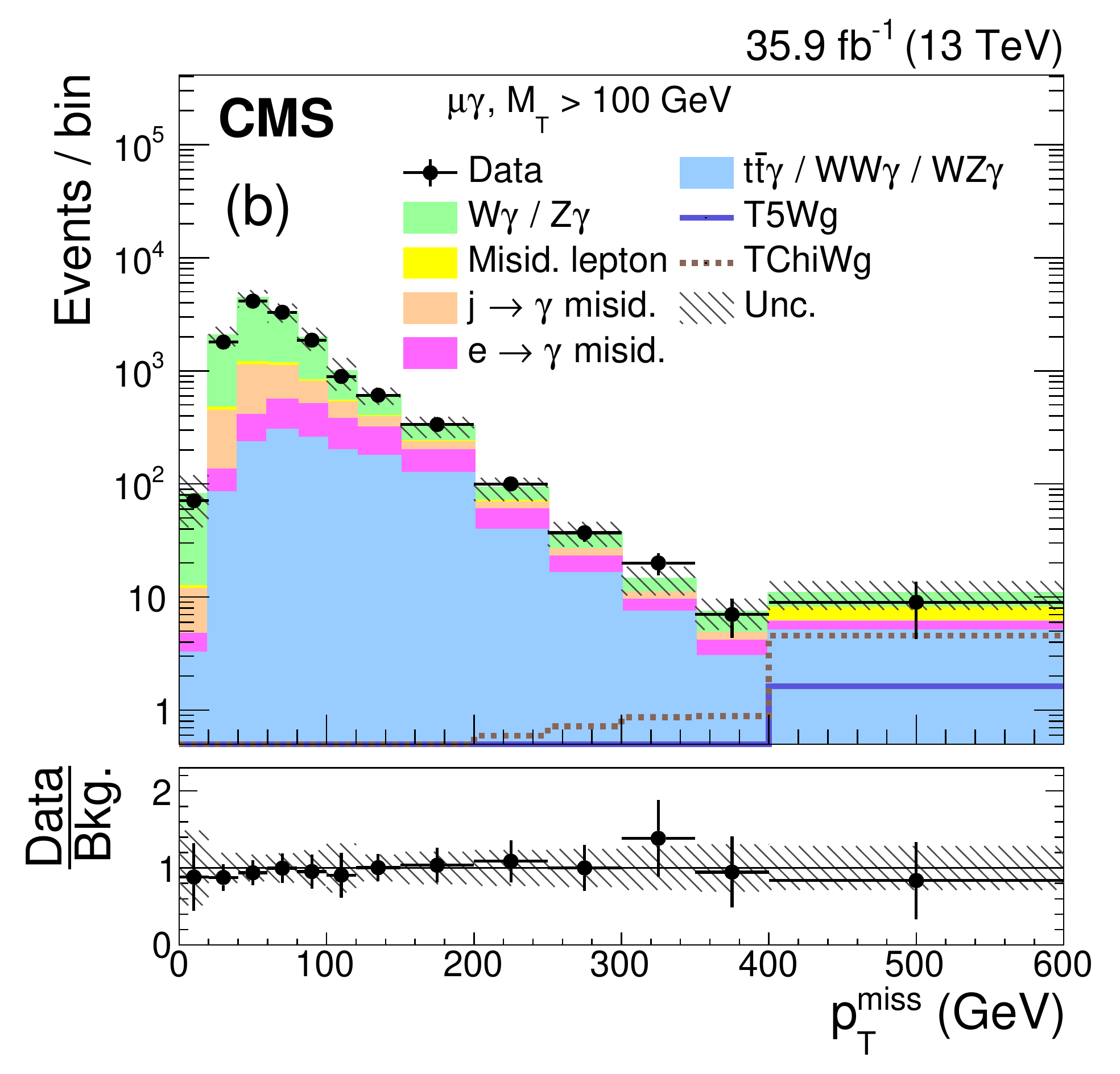} \\
\includegraphics[width=0.39\textwidth]{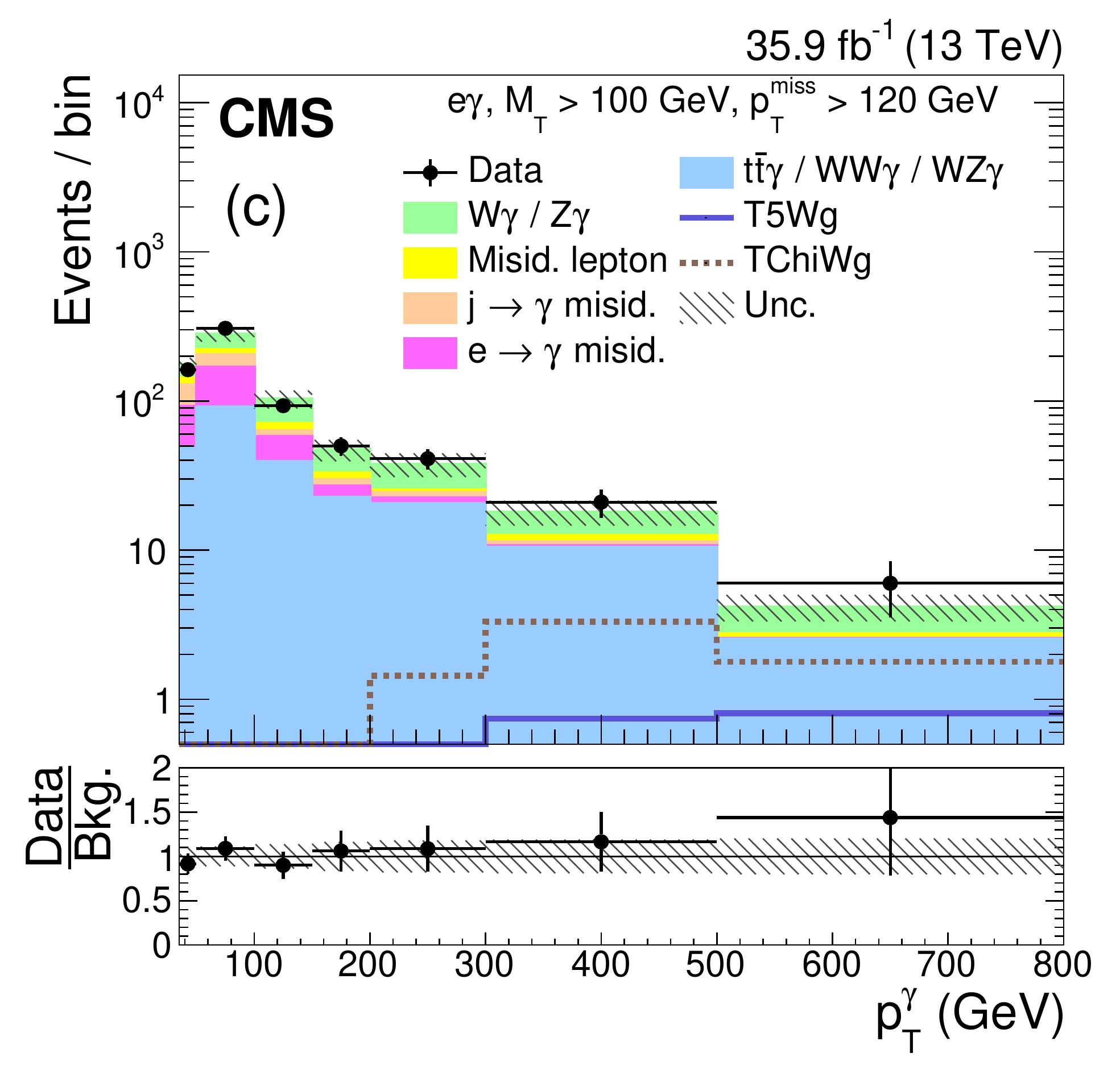}
\includegraphics[width=0.39\textwidth]{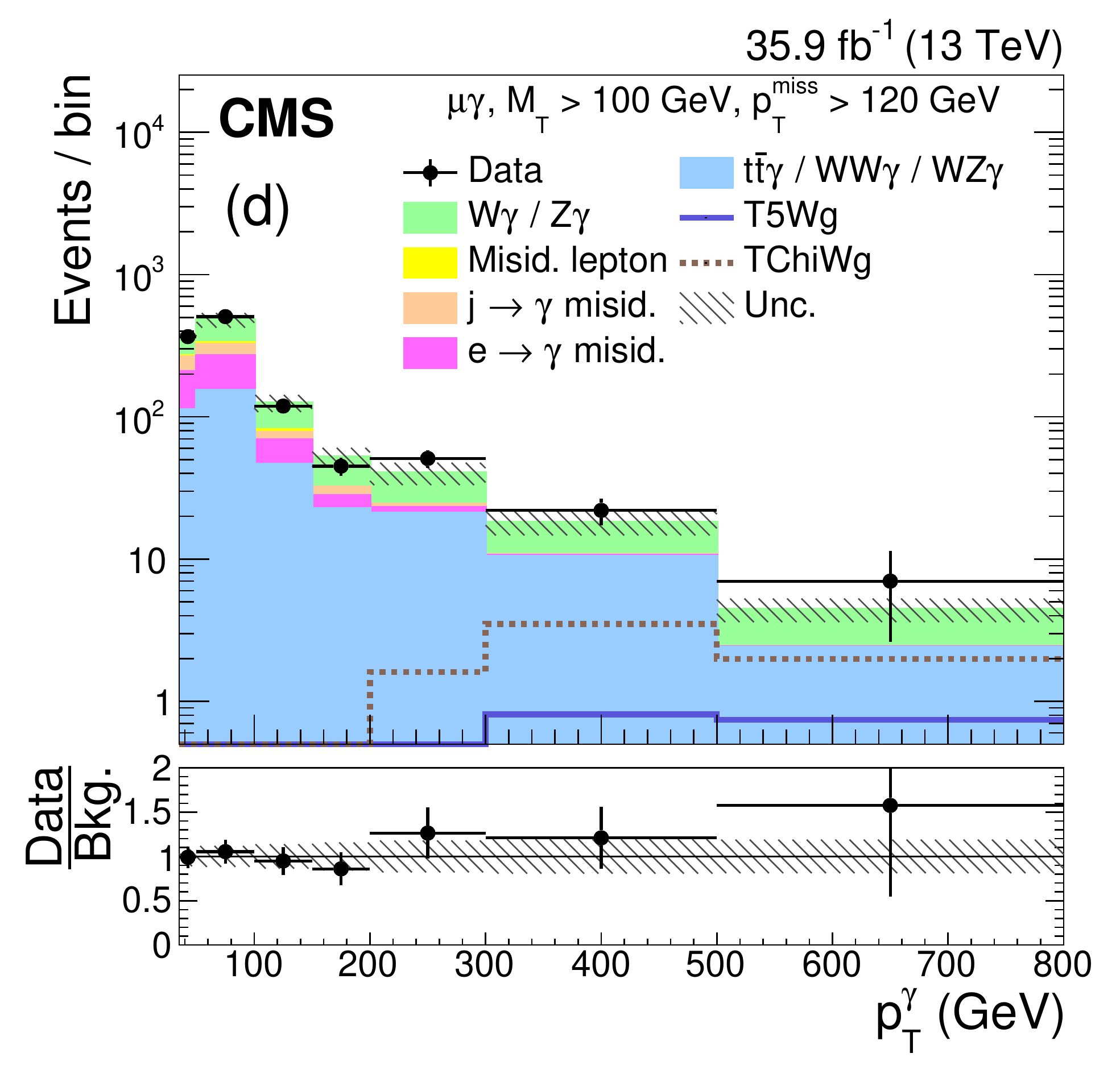} \\
\includegraphics[width=0.39\textwidth]{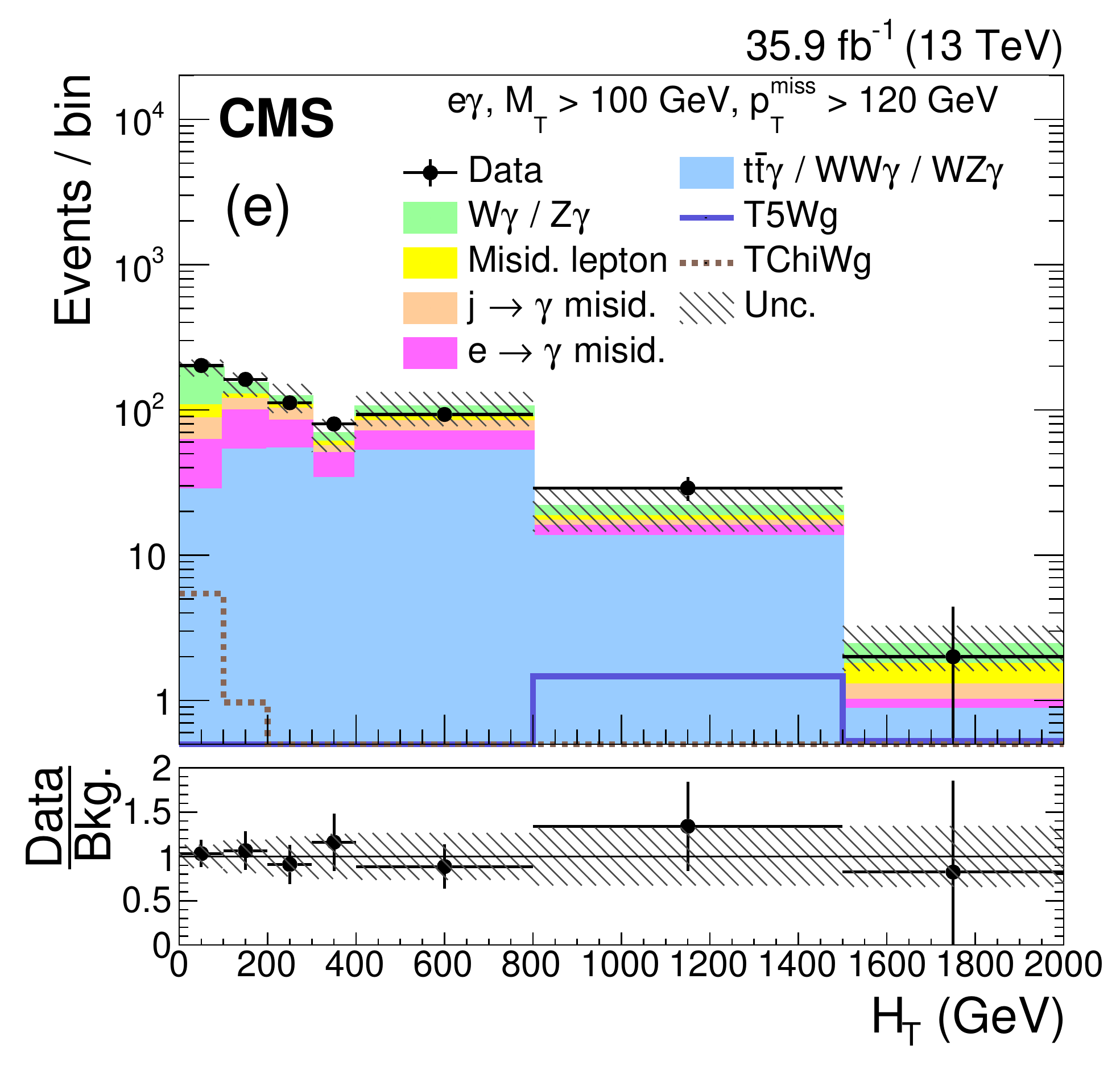}
\includegraphics[width=0.39\textwidth]{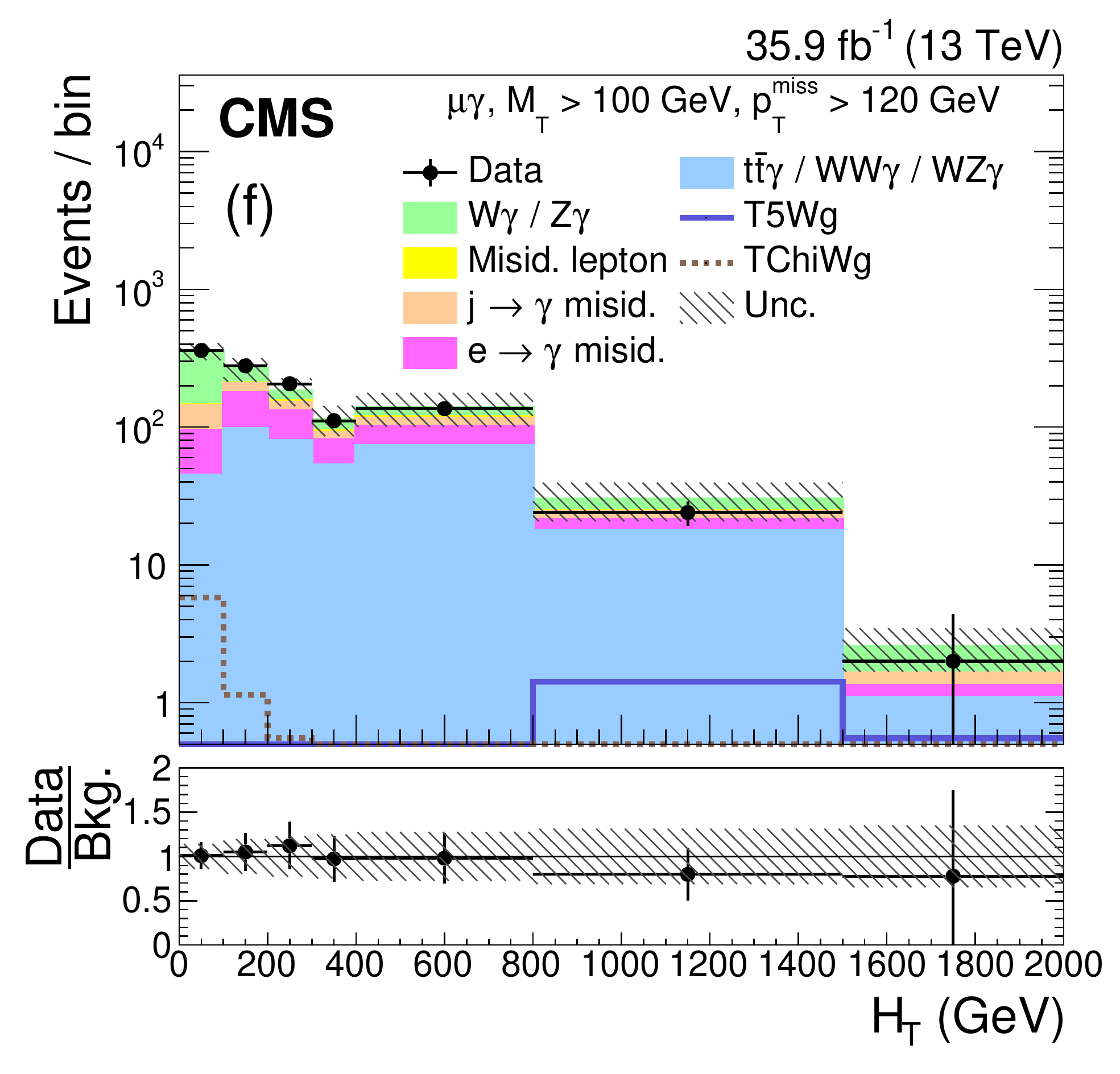}

\caption{Distributions of \ptmiss (a, b), $\pt^{\Pgg}$ (c, d), and \HT
  (e, f) from data (points) and estimated SM predictions (stacked
  histograms) for the {\re\Pgg} (left) and $\mu${\Pgg} (right)
  channels. Simulated signal distributions from the TChiWg model
  (dotted) with $m_{\PSGcz/\PSGcpm} = 800\GeV$ and the T5Wg model
  (solid) with $m_{\PSg} = 1700\GeV$ are overlaid. The \ptmiss
  distribution includes all events with $\mT > 100\GeV$, while the
  $\pt^{\Pgg}$ and \HT distributions only include events with $\mT >
  100\GeV$ and $\ptmiss > 120\GeV$. The vertical bars on the points give
  the statistical uncertainty in the data and the horizontal bars show
  the bin widths. The hatched area represents the quadratic sum of the
  statistical and systematic uncertainties in the simulated
  background. The lower panels display the ratio of the data to the
  total background prediction. }
    \label{fig:unblinding}
\end{figure}

Figure~\ref{fig:unblinding} shows the \ptmiss, $\pt^{\Pgg}$, and {\HT}
distributions of the observed data and predicted background, together
with the systematic uncertainties in the background prediction. The
\ptmiss distribution includes all events with $\mT > 100\GeV$, while the
$\pt^{\Pgg}$ and \HT distributions only include events in the signal
region. Two simulated signal distributions, one from the TChiWg
simplified model with an NLSP mass of 800\GeV, and the other from the
T5Wg model with an NLSP mass of 1000\GeV and a gluino mass of 1700\GeV,
are also overlaid. The data are compatible with the estimated SM
backgrounds within the uncertainties.

\begin{figure}[hbtp]
    \centering
    \includegraphics[width=0.8\textwidth]{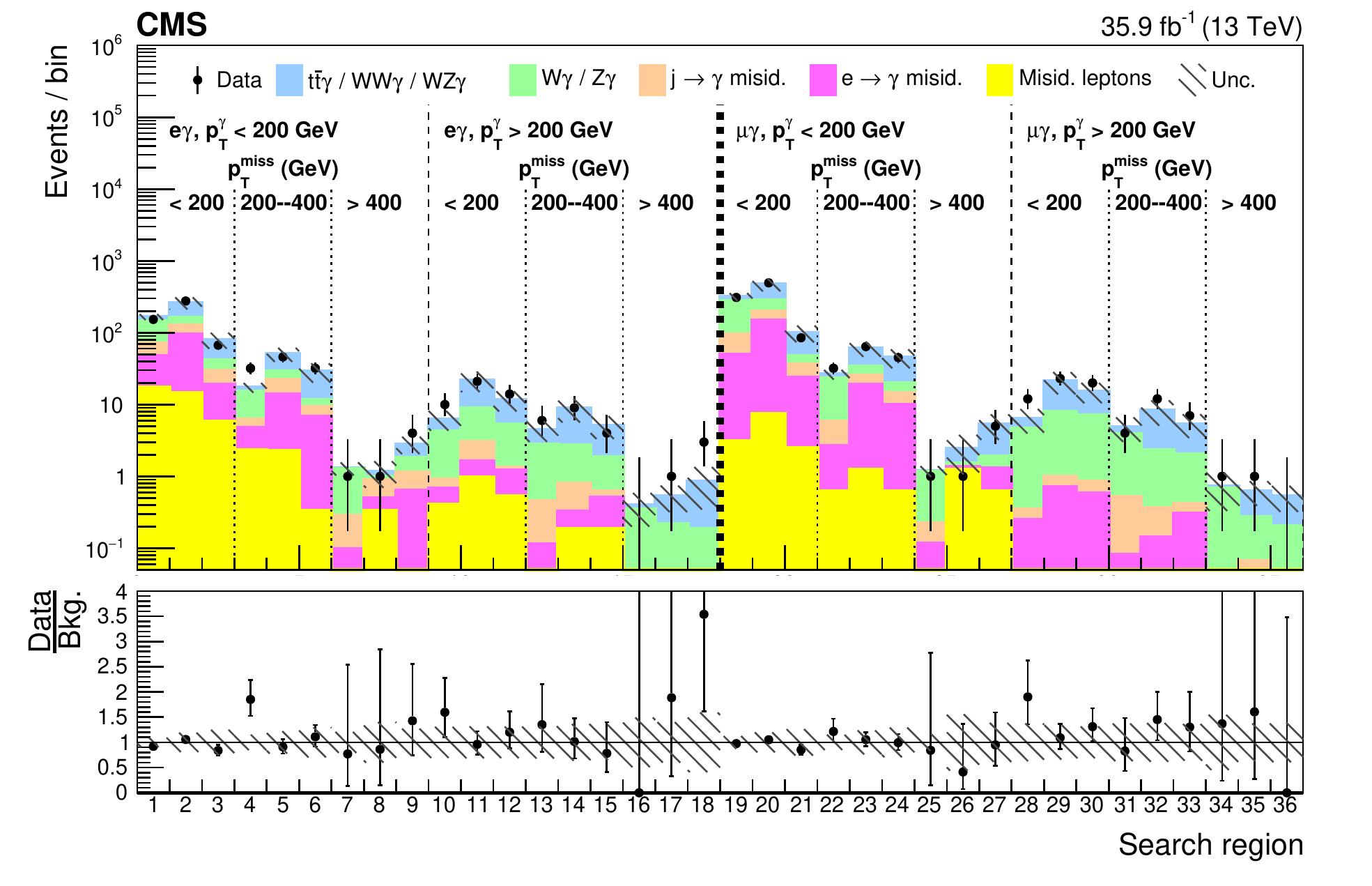}
    \caption{The number of data events (points) and predicted background
      events (shaded histograms) for the 18 search regions in \ptmiss,
      \HT, and $\pt^{\Pgg}$ (separated by dashed vertical lines) in the
      {\Pe\Pgg} (regions 1--18) and the $\mu${\Pgg} (regions 19--36)
      channels. For each \ptmiss range, the first, second, and last bins
      correspond to the \HT regions 0--100, 100--400, and $>400\GeV$,
      respectively. The lower panel displays the ratio of the data to
      the background predictions.  The vertical bars on the points show
      the statistical uncertainty in the data, and the hatched areas
      give the quadrature sum of the statistical and systematic
      uncertainties in the simulated background.}
    \label{fig:signal}
\end{figure}

To improve the sensitivity for different SUSY scenarios, the signal
region for each lepton channel is further divided into 18 search
regions: three bins of \ptmiss (120--200, 200--400, and ${>}400\GeV$) in
each of three {\HT} ranges (0--100, 100--400, and ${>}400\GeV$), and two
ranges of photon \pt (35--200 and ${>}200\GeV$).  The
misidentified-photon and misidentified-lepton control samples are also
divided into respective search regions.  Figure~\ref{fig:signal} gives
the event yields from data and the estimated total background in each of
the search regions for the $\Pe\Pgg$ (left part) and $\mu\Pgg$ (right
part) channels.  The observed data are consistent with the background
predictions in all the search regions.  The largest difference is in the
fourth bin of the $\Pe\Pgg$ channel, which has an excess over the
background prediction of 2.3 standard deviations.  In the corresponding
search regions of the $\mu${\Pgg} channel, the data are compatible with
the SM background predictions. Thus, we conclude that no significant
excess of events beyond the SM expectation is observed.

\section{Interpretation}
\label{sec:interpretation}

The results are interpreted in the context of upper limits on the cross
sections of the three simplified SUSY models introduced in
section~\ref{sec:introduction}.  For each mass point of the signal
models, a 95\% confidence level (\CL) upper limit on the signal
production cross section is obtained by calculating \CLs
limits~\cite{Junk:1999kv,Read:2002hq,CMS-NOTE-2011-005} using the
profile likelihood as a test statistic and asymptotic
formulas~\cite{Cowan:2010js}.  The SM background prediction, signal
expectation, and observed number of events in each signal search region
of the \Pe\Pgg\ and \Pgm\Pgg\ channels defined above are combined into
one statistical interpretation, and studied as a multichannel counting
experiment.

\begin{figure}[tbp]
  \centering
    \includegraphics[width=0.6\textwidth]{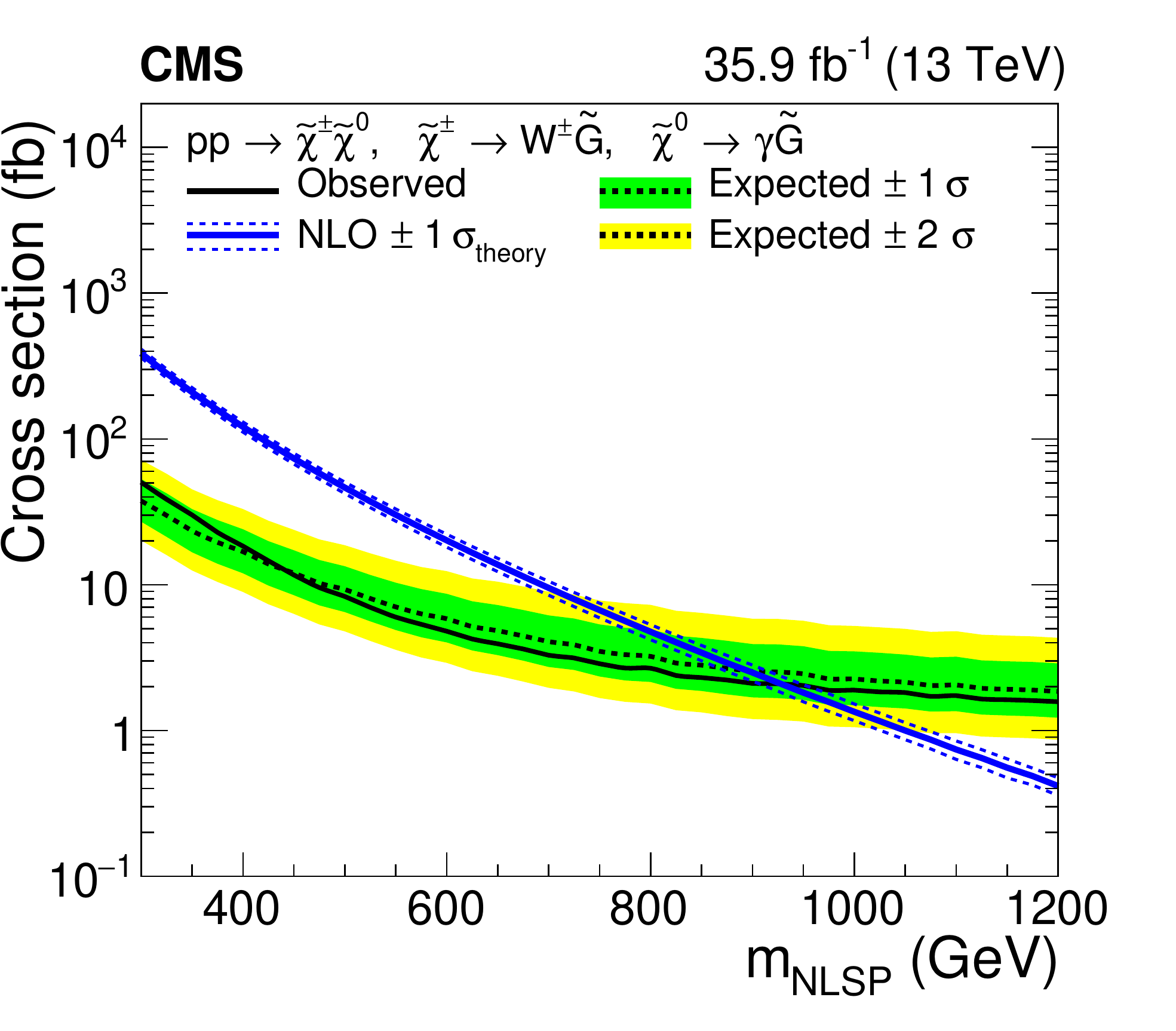}
    \caption{The observed (solid line) and expected (dashed line) 95\%
      {\CL} upper limits on the production cross sections for the TChiWg
      simplified model, together with the NLO theoretical cross sections
      as a function of the NLSP mass.  The inner (darker) band and outer
      (lighter) band around the expected upper limits indicate the
      regions containing 68 and 95\%, respectively, of the distribution
      of limits expected under the background-only hypothesis. The
      dotted lines around the theoretical cross section gives the $\pm1$
      standard deviation uncertainty in the cross section.  }
    \label{fig:tchiwg}
\end{figure}

Figure \ref{fig:tchiwg} shows the observed and expected 95\% \CL upper
limits on the cross section for the TChiWg model as a function of the
NLSP mass, together with the theoretical cross section for
$\PSGcz\PSGcpm$ pair production. The TChiWg model is based on the direct
production of \PSGcpm\ and \PSGcz, in which their decays are restricted
to $\PW^{\pm}\PXXSG$ and $\Pgg\PXXSG$, respectively. The gravitino
\PXXSG\ is modeled as nearly massless. Assuming a 100\% branching
fraction for $\PSGcz\to\Pgg\PXXSG$, this search excludes NLSP masses up
to 930\GeV at the 95\% \CL.

\begin{figure}[tbp]
  \centering
    \includegraphics[width=0.49\textwidth]{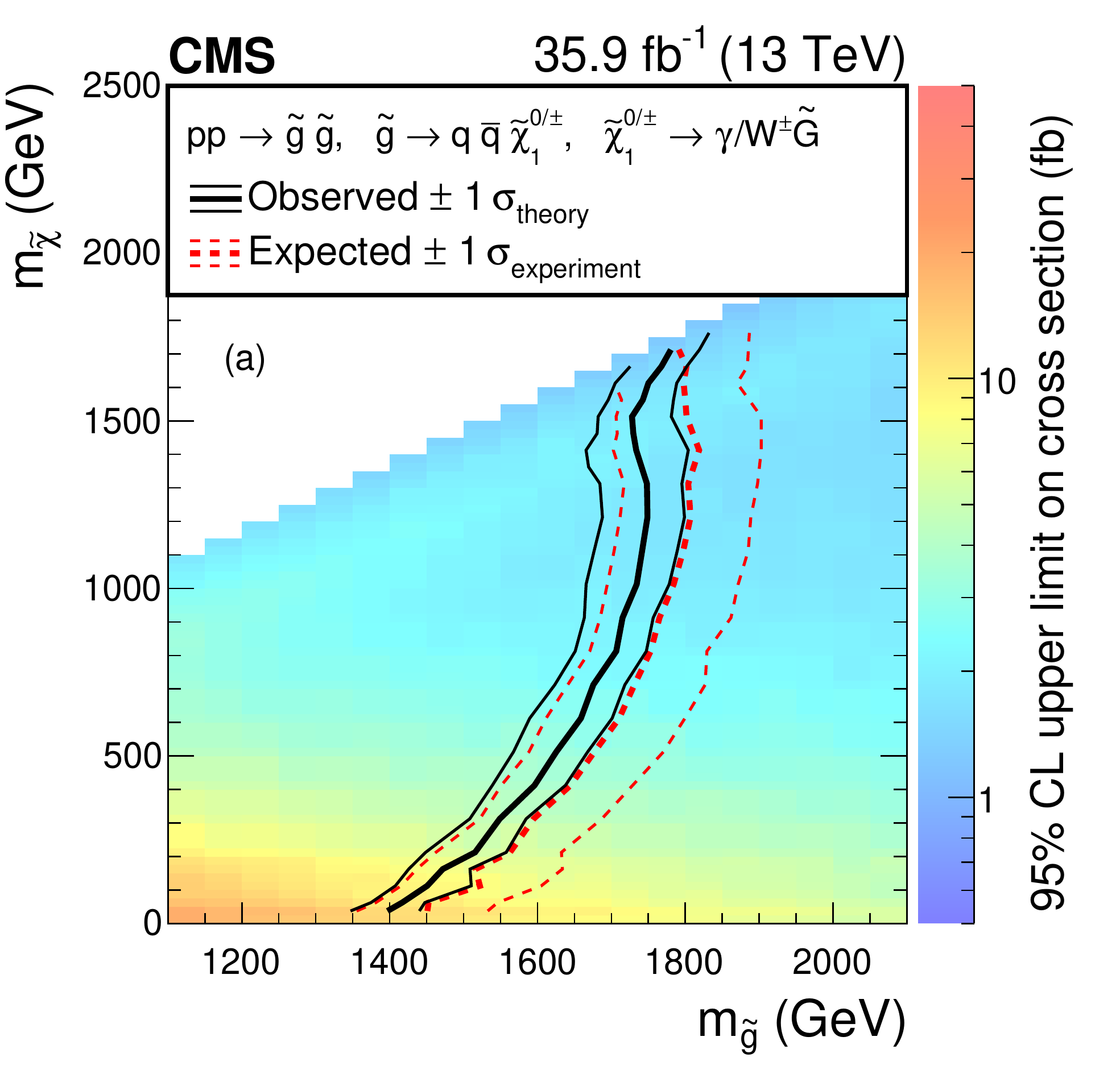}
    \includegraphics[width=0.49\textwidth]{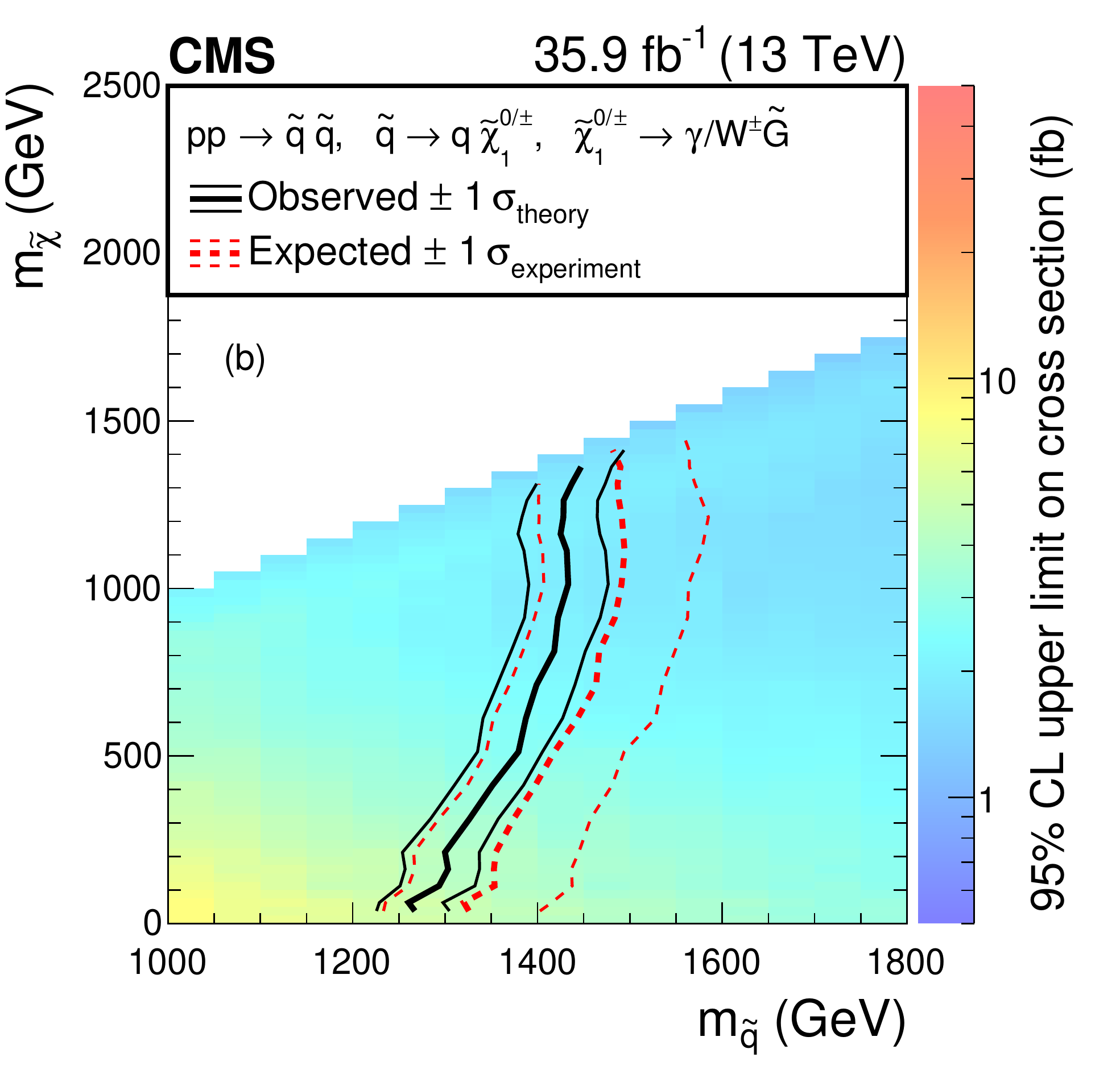}
    \caption{ The observed (solid line) and expected (dashed line) 95\%
      {\CL} exclusion contours for (a) $m_{\PSg}$ versus $m_{\PSGc}$ and
      (b) $m_{\PSq}$ versus $m_{\PSGc}$ (regions to the left of the
      curves are excluded), and the 95\% {\CL} upper limits on the pair
      production cross sections for (a) {\PSg\PSg} in the T5Wg and (b)
      {\PSq\PSq} in the T6Wg simplified models (use the scales to the
      right of the plots).  The upper limits on the cross sections
      assume a 50\% branching fraction for $\PSg\to
      \Pq\Paq\PSGcz/\PSGcpm$ and $\PSq\to \Pq\PSGcz/\PSGcpm$.  The bands
      around the observed and expected exclusion contours indicate the
      $\pm1$ standard deviation range when including the experimental
      and theoretical uncertainties, respectively.}

    \label{fig:t5wg}
\end{figure}

In figure~\ref{fig:t5wg}, we present the cross section 95\% {\CL} upper
limits and mass exclusion contours for the T5Wg and T6Wg simplified
models.  The production cross section of the T5Wg (T6Wg) model is
determined solely by $m_{\PSg}$ ($m_{\PSq}$). Nevertheless, the
$m_{\PSg/\PSq} - m_{\PSGc}$ mass difference affects the \HT and \ptmiss
spectra, resulting in nontrivial exclusion-limit contours in the
$(m_{\PSg/\PSq}, m_{\PSGc})$ mass plane. The branching fractions for
$\PSg\to \Pq\Paq\PSGcz/\PSGcpm$ and $\PSq\to \Pq\PSGcz/\PSGcpm$ are
assumed to be 50\%. For large $\PSGcz/\PSGcpm$ masses, gluino (squark)
masses are excluded at 95\% {\CL} up to 1.75 (1.43)\TeV in the T5Wg
(T6Wg) scenarios.

\section{Summary}

A search for supersymmetry with general gauge mediation in events with a
photon, an electron or muon, and large missing transverse momentum has
been presented. This analysis is based on a sample of proton-proton
collisions at $\sqrt{s} = 13\TeV$, corresponding to an integrated
luminosity of 35.9\fbinv recorded by the CMS experiment in 2016. The
data are examined in bins of the photon transverse energy, the magnitude
of the missing transverse momentum, and the scalar sum of jet
energies. The standard model background is evaluated primarily using
control samples in the data, with simulation used to evaluate
backgrounds from electroweak processes.  The data are found to agree
with the standard model expectation, without significant excess in the
search region.  The results of the search are interpreted as 95\%
confidence level upper limits on the production cross sections of
supersymmetric particles in the context of simplified
models~\cite{Chatrchyan:2013sza} motivated by gauge-mediated
supersymmetry breaking. For strong production models, such as the T5Wg
simplified model of gluino pair production and the T6Wg model of squark
pair production, this search excludes gluinos (squarks) with masses up
to 1.75 (1.43)\TeV in the T5Wg (T6Wg) scenarios.  The TChiWg simplified
model, based on direct electroweak production of a neutralino and
chargino, is excluded for next-to-lightest supersymmetric particle
masses below 930\GeV, extending the current best limit by about
150\GeV~\cite{2018118}.

\begin{acknowledgments}
We congratulate our colleagues in the CERN accelerator departments for
the excellent performance of the LHC and thank the technical and
administrative staffs at CERN and at other CMS institutes for their
contributions to the success of the CMS effort. In addition, we
gratefully acknowledge the computing centers and personnel of the
Worldwide LHC Computing Grid for delivering so effectively the computing
infrastructure essential to our analyses. Finally, we acknowledge the
enduring support for the construction and operation of the LHC and the
CMS detector provided by the following funding agencies: BMWFW and FWF
(Austria); FNRS and FWO (Belgium); CNPq, CAPES, FAPERJ, and FAPESP
(Brazil); MES (Bulgaria); CERN; CAS, MoST, and NSFC (China); COLCIENCIAS
(Colombia); MSES and CSF (Croatia); RPF (Cyprus); SENESCYT (Ecuador);
MoER, ERC IUT, and ERDF (Estonia); Academy of Finland, MEC, and HIP
(Finland); CEA and CNRS/IN2P3 (France); BMBF, DFG, and HGF (Germany);
GSRT (Greece); OTKA and NIH (Hungary); DAE and DST (India); IPM (Iran);
SFI (Ireland); INFN (Italy); MSIP and NRF (Republic of Korea); LAS
(Lithuania); MOE and UM (Malaysia); BUAP, CINVESTAV, CONACYT, LNS, SEP,
and UASLP-FAI (Mexico); MBIE (New Zealand); PAEC (Pakistan); MSHE and
NSC (Poland); FCT (Portugal); JINR (Dubna); MON, RosAtom, RAS, RFBR and
RAEP (Russia); MESTD (Serbia); SEIDI, CPAN, PCTI and FEDER (Spain);
Swiss Funding Agencies (Switzerland); MST (Taipei); ThEPCenter, IPST,
STAR, and NSTDA (Thailand); TUBITAK and TAEK (Turkey); NASU and SFFR
(Ukraine); STFC (United Kingdom); DOE and NSF (USA).

 \hyphenation{Rachada-pisek} Individuals have received support from the
 Marie-Curie program and the European Research Council and Horizon
 2020 Grant, contract No. 675440 (European Union); the Leventis
 Foundation; the A. P. Sloan Foundation; the Alexander von Humboldt
 Foundation; the Belgian Federal Science Policy Office; the Fonds pour
 la Formation \`a la Recherche dans l'Industrie et dans l'Agriculture
 (FRIA-Belgium); the Agentschap voor Innovatie door Wetenschap en
 Technologie (IWT-Belgium); the F.R.S.-FNRS and FWO (Belgium) under the
 ``Excellence of Science - EOS'' - be.h project n. 30820817; the Ministry
 of Education, Youth and Sports (MEYS) of the Czech Republic; the
 Lend\"ulet (``Momentum'') Program and the J\'anos Bolyai Research
 Scholarship of the Hungarian Academy of Sciences, the New National
 Excellence Program \'UNKP, the NKFIA research grants 123842, 123959,
 124845, 124850 and 125105 (Hungary); the Council of Science and
 Industrial Research, India; the HOMING PLUS program of the Foundation
 for Polish Science, cofinanced from European Union, Regional
 Development Fund, the Mobility Plus program of the Ministry of
 Science and Higher Education, the National Science Center (Poland),
 contracts Harmonia 2014/14/M/ST2/00428, Opus 2014/13/B/ST2/02543,
 2014/15/B/ST2/03998, and 2015/19/B/ST2/02861, Sonata-bis
 2012/07/E/ST2/01406; the National Priorities Research Program by Qatar
 National Research Fund; the Programa Estatal de Fomento de la
 Investigaci{\'o}n Cient{\'i}fica y T{\'e}cnica de Excelencia Mar\'{\i}a
 de Maeztu, grant MDM-2015-0509 and the Programa Severo Ochoa del
 Principado de Asturias; the Thalis and Aristeia programs cofinanced
 by EU-ESF and the Greek NSRF; the Rachadapisek Sompot Fund for
 Postdoctoral Fellowship, Chulalongkorn University and the Chulalongkorn
 Academic into Its 2nd Century Project Advancement Project (Thailand);
 the Welch Foundation, contract C-1845; and the Weston Havens Foundation
 (USA).
\end{acknowledgments}

\bibliography{auto_generated}
\cleardoublepage \appendix\section{The CMS Collaboration \label{app:collab}}\begin{sloppypar}\hyphenpenalty=5000\widowpenalty=500\clubpenalty=5000\vskip\cmsinstskip
\textbf{Yerevan Physics Institute, Yerevan, Armenia}\\*[0pt]
A.M.~Sirunyan, A.~Tumasyan
\vskip\cmsinstskip
\textbf{Institut f\"{u}r Hochenergiephysik, Wien, Austria}\\*[0pt]
W.~Adam, F.~Ambrogi, E.~Asilar, T.~Bergauer, J.~Brandstetter, M.~Dragicevic, J.~Er\"{o}, A.~Escalante~Del~Valle, M.~Flechl, R.~Fr\"{u}hwirth\cmsAuthorMark{1}, V.M.~Ghete, J.~Hrubec, M.~Jeitler\cmsAuthorMark{1}, N.~Krammer, I.~Kr\"{a}tschmer, D.~Liko, T.~Madlener, I.~Mikulec, N.~Rad, H.~Rohringer, J.~Schieck\cmsAuthorMark{1}, R.~Sch\"{o}fbeck, M.~Spanring, D.~Spitzbart, A.~Taurok, W.~Waltenberger, J.~Wittmann, C.-E.~Wulz\cmsAuthorMark{1}, M.~Zarucki
% ...
\vskip\cmsinstskip
% ...
1:  Also at Vienna University of Technology, Vienna, Austria\\
% ...
\end{sloppypar}
\end{document}